\documentstyle[12pt,epsfig]{article}
\newlength{\dinwidth}
\newlength{\dinmargin}
\newcommand{\ba}{\begin{array}}
\newcommand{\ea}{\end{array}}
\newcommand{\beq}{\begin{equation}}
\newcommand{\eeq}{\end{equation}}
\newcommand{\bea}{\begin{eqnarray}}
\newcommand{\eea}{\end{eqnarray}}

\def\bce{\begin{center}}
\def\ece{\end{center}}

\def\nonu{\nonumber}

\def\pa{\partial}

\def\be{\beta}
\def\ga{\gamma}

\def\de{\delta}

\def\eps6{{\displaystyle \mathop{\epsilon}^{6}}{}}

\def\nab6{{\displaystyle \mathop{\nabla}^{6}}{}}

\def\ba{\begin{array}}
\def\ea{\end{array}}
\def\beq{\begin{equation}}
\def\eeq{\end{equation}}
\def\be{\begin{equation}}
\def\ee{\end{equation}}

\def\d{\partial}

\def\eps{\epsilon}

\def\d{{\rm d}}

\def\ba{\begin{array}}
\def\ea{\end{array}}
\def\beq{\begin{equation}}
\def\eeq{\end{equation}}
\def\be{\begin{equation}}
\def\ee{\end{equation}}

\def\d{\partial}

\def\eps{\epsilon}

\def\d{{\rm d}}

\newcommand{\bean}{\begin{eqnarray*}}
\newcommand{\eean}{\end{eqnarray*}}

\begin{document}
\thispagestyle{empty} \addtocounter{page}{-1}
\begin{flushright}
{\tt hep-th/0407009}\\
\end{flushright}

\vspace*{1.3cm} 
\centerline{ \Large \bf Mirror Symmetry of Calabi-Yau 
Supermanifolds }
\vspace*{1.5cm}
\centerline{{\bf Changhyun Ahn
}} 
\vspace*{1.0cm} 
 \centerline{\it  School of Natural Sciences,
Institute for Advanced Study,
Olden Lane, Princeton NJ 08540, USA}
\centerline{\it Department of Physics,
Kyungpook National University, Taegu 702-701, Korea}
\vspace*{0.8cm} 
\centerline{\tt
ahn@ias.edu}  
\vskip2cm

\centerline{\bf Abstract}
\vspace*{0.5cm}

We study  super Landau-Ginzburg mirrors of the 
weighted projective superspace ${\bf WCP }^{3|2}$
which is a Calabi-Yau supermanifold and appeared in 
hep-th/0312171(Witten) in the topological B-model.
One of them is 
an elliptic fibration 
over the complex plane 
whose coordinate is given in terms of two bosonic and
two fermionic variables  as well as Kahler parameter
of  ${\bf WCP }^{3|2}$.  
The other is some patch of
a degree 3 Calabi-Yau hypersurface in ${\bf CP}^2$
fibered by the complex plane whose coordinate depends on 
both 
above four variables and Kahler parameter 
but its dependence behaves quite differently.

\baselineskip=18pt
\newpage
\renewcommand{\theequation}{\arabic{section}\mbox{.}\arabic{equation}}

\section{Introduction}
\setcounter{equation}{0}

\indent

From the equivalence between 
the perturbative expansion of ${\cal N}=4$ super Yang-Mills
theory and the D-instanton expansion of a  topological B-model 
on the Calabi-Yau {\it supermanifold} 
${\bf CP}^{3|4}(1,1,1,1|1,1,1,1)$,
the planar amplitudes of ${\cal N}=4$ super Yang-Mills 
theory are 
supported to holomorphic curves \cite{Witten}.
This target space has four bosonic homogeneous coordinates
of weight 1 and four fermionic homogeneous coordinates
of weight 1.
The idea of Calabi-Yau supermanifold was found in \cite{Sethi} 
to resolve some issues in the mirror symmetry (for the review of 
this, see \cite{HoriVafa,HIV}) 
by extending 
the space of bosonic Calabi-Yau manifold to the space of Calabi-Yau 
supermanifold. Moreover in \cite{Schwarz}, it was shown that
certain Calabi-Yau space and Calabi-Yau superspace in the 
topological A-model are equivalent
to each other.

The topological B-model on ${\bf CP}^{3|4}$ is mapped to 
A-model on ${\bf CP}^{3|4}$ through S-duality \cite{NV} and by 
mirror symmetry it was {\it conjectured} 
that the topological A-model on
this Calabi-Yau supermanifold 
is equivalent to the topological B-model on
a quadric in ${\bf CP}^{3|3} \times {\bf CP}^{3|3}$ \cite{Witten1978}. 
The complex
supermanifold ${\bf CP}^{3|3}$ has four bosonic homogeneous coordinates
of weight 1 and three fermionic homogeneous coordinates
of weight 1 and is not Calabi-Yau supermanifold. However, a quadric
is a Calabi-Yau supermanifold. Moreover,
it was {\it shown} 
that the topological A-model on the ${\bf CP}^{3|4}$
is mirror to the B-model on the quadric in
${\bf CP}^{3|3} \times {\bf CP}^{3|3}$
in the particular limit of Kahler parameter of ${\bf CP}^{3|4}$ 
\cite{AV}.
Furthermore, the topological A-model on a quadric in ${\bf CP}^{3|3}
\times {\bf CP}^{3|3}$ is mirror to the B-model on ${\bf CP}^{3|4}$
in a certain limit of one of the two Kahler parameters of the quadric
and it has been checked the previous conjecture by \cite{NV} in \cite{Kumar}. 

In \cite{Witten}, 
it was suggested that a possible generalization of a target 
space can be done by giving a different weight to the fermionic 
coordinates without changing the bosonic manifold ${\bf CP}^3$.  
To construct Calabi-Yau supermanifold, the sum of bosonic
weights equals that of fermionic weights. By allowing the 
fermionic weights to be positive and odd, one can write this supermanifold
as a weighted projective superspace
${\bf WCP }^{3|2}(1,1,1,1|1,3)$ uniquely.  
The central charge $\hat{c}$ for a conformal supersymmetric 
sigma  model is given by $\hat{c}=3-2=1$ which is defined to be the 
net complex superdimension. The dimension of bosonic manifold 
is equal to 3 while the dimension of fermionic
manifold is equal to 2.

In this note, we generalize the mirror geometry \cite{HoriVafa} 
from the linear sigma model
description \cite{Witten1993} in the context of bosonic Calabi-Yau
manifold to the case in the above Calabi-Yau {\it supermanifold}
${\bf WCP}^{3|2}$.
The idea of \cite{AV} is to introduce 
two additional fermions as well 
as a bosonic superfield when 
we take T-dualization for the phase of each fermion superfield.
In section 2,
starting with the super Landau-Ginzburg(LG) 
B-model mirror combined with the
prescription of \cite{AV}(the concept of T-duality for a fermion 
field), we compute some path integrations over 
dual bosonic and fermionic superfields. 
These manipulations are extremely `formal' with target 
supermanifold 
in the sense that
the field theory is nonunitary and usually 
does not have a good vacuum. 
With the same spirit of \cite{NV}, 
our starting point is to take 
A-model on  ${\bf WCP}^{3|2}$ which is mapped 
to B-model on the same space by S-duality and study 
the B-model mirror of the topological A-model on
${\bf WCP}^{3|2}$.
We use this conjecture and some formal manipulations to see 
where they lead to 
some
hypersurface equation satisfied by mirror 
Calabi-Yau supermanifold and 
its mirror geometry.  
In section 3, after the summary of the paper, we make some 
remarks and future direction.

\section{Mirror of ${\bf WCP }^{3|2}(1,1,1,1|1,3)$}
\setcounter{equation}{0}

\indent

The  weighted projective superspace
${\bf WCP }^{3|2}(1,1,1,1|1,3)$ has 
an extension of a linear sigma model \cite{Witten1993}
description in terms of four bosonic homogeneous 
coordinates $Z^I(I=1,2,3,4)$ of weight 1 and two fermionic 
homogeneous coordinates $\psi, \chi$ of weights 1 and 3, 
respectively.
Since the sum of bosonic weights equals the sum of fermionic 
weights,  ${\bf WCP }^{3|2}(1,1,1,1|1,3)$ is a Calabi-Yau 
supermanifold \cite{Witten}.  
One can define a topological B-model with this target space 
\cite{Witten,PW}
and this
supermanifold admits ${\cal N}=1$ superconformal symmetry 
acting on both $Z^I$ and $\psi$.
The geometry of the linear sigma model 
with a given complexified Kahler class 
parameter $t$  can be analyzed by 
solving the following D-term constraint
\bea
 \sum_{I=1}^4 |Z^I|^2  + |\psi|^2+ 3 |\chi|^2 =
\mbox{Re} \; t
\nonu
\eea 
and dividing the gauge group $U(1)$.
The vacuum structure depends on the signature of 
Kahler parameter
$\mbox{Re} \;t$.

Under the T-duality, the bosonic superfields $Z^I$ 
of the linear sigma model 
are replaced by a dual cylinder-valued superfield $Y_I$,
as usual, 
and the fermionic superfields
$\psi$ and $\chi$ are dualized to 
$(X_1, \eta_1, \chi_1)$ and $(X_2, \eta_2, \chi_2)$
respectively \cite{AV}.  For each fermion field, there are
a bosonic mirror 
as well as two additional fermions in order to preserve 
the central charge(that is, there is 
only one net fermionic dimension for the 
mirror). 
Here $X_1$ and $X_2$ are bosonic superfields while
$\eta_i$ and $\chi_i$ where $i=1,2$ 
are {\it fermionic} superfields. 
Then the super Landau Ginzburg B-model mirror of  
${\bf WCP }^{3|2}(1,1,1,1|1,3)$ is
given by the path integral for the holomorphic sector
\cite{AV}
\bea
\int  \prod_{I=1}^{4} d Y_I  
\prod_{J=1}^{2} d X_J d \eta_J \d \chi_J
\de \left(\sum_{I=1}^{4} Y_I- X_1-3X_2 -t  \right)  
\exp \left[\sum_{I=1}^{4} e^{-Y_I}  + \sum_{J=1}^{2} 
e^{-X_J} \left(1+ \eta_J \chi_J \right) \right].
\nonu
\eea
The superpotential of the mirror theory
can be read off from the above exponent.
The super LG model has 5 bosonic and 4 fermionic 
degrees of freedom(a mirror manifold has the same superdimension 
$5-4=1$ as the original one, by mirror symmetry). 
Note that there exists a single delta function
constraint where $\mbox{Re} \;Y_I =|Z^I|^2, \mbox{Re} 
\;X_1 =-|\psi|^2$ and $\mbox{Re} \;X_2 = -3|\chi|^2$.  

We want to rewrite the above as a sigma model on a 
super Calabi-Yau hypersurface.
To describe a mirror super Calabi-Yau interpretation for
this LG, we should manipulate 
the integrations over some superfields and 
successive superfield redefinitions.
There exist two possibilities to carry out the integrations 
over the fermion superfields. 
In subsection 
\ref{X1},
we first consider 
the case where the $X_1$ is given in terms of  
other variables and a complexified Kahler parameter
$t$
using the delta function constraint above. 
The final integral has 3 bosonic and 2 fermionic 
degrees of freedom.
In subsection \ref{X2},
we will come to the second case where we 
integrate out the $X_2$(as well as two fermionic 
superfields).

\subsection{
Integration over $X_1$}
\label{X1}

\indent

To describe a mirror super Calabi-Yau interpretation for this
super LG, we integrate out 
some superfields with appropriate successive superfield 
redefinitions.  
In terms of $Y_I$ and $X_2$, the 
delta function 
allows us to write $X_1$ as follows: 
\bea
X_1= -3X_2 -t+\sum_{I=1}^{4} Y_I. 
\label{delta}
\eea
We first integrate out the fermions $\eta_1$ and $\chi_1$
and solve the delta function constraint for $X_1$.
By computing the integrations over $\eta_1, \chi_1$ and $X_1$,
one gets the following B-model integral with 5 bosons and 2
fermions
\bea 
\int
\left( \prod_{I=1}^{4} d Y_I \right)
d X_2 d \eta_2 d \chi_2
 \;\; e^{3X_2-\sum_{I=1}^{4} Y_I} 
 \exp \left[   
\sum_{I=1}^{4} e^{-Y_I}  +  
e^{-X_2} \left(1+ \eta_2 \chi_2\right)
+ e^{t + 
3X_2 - \sum_{I=1}^{4} Y_I }  \right].
\nonu
\eea
The nontrivial factors in the measure come from the 
integrating out the fermion superfields and we  
ignore an irrelevant 
normalization $e^t$. In the last term inside the exponent, we
replaced $X_1$ with $X_2, Y_I$ and $t$, according to the delta 
function constraint (\ref{delta}). 
Now let us introduce the new $\bf C$-valued bosonic superfields 
$x_2$ and $y_I$(which are good variables) as follows:
\bea
e^{-X_2} \equiv    x_2, \qquad e^{-Y_I} \equiv y_I, \qquad
I=1,2,3,4.
\nonu
\eea

Then in terms of these new fields \footnote{One can also introduce
the new variables $Y_i=\hat{Y}_i +Y_1$ where $i=2,3,4$  
and $X_2=\hat{X}_2 +Y_1$
and in this case $e^{-Y_1}$ plays the role of a Lagrange multiplier
and
the superpotential has an overall factor $e^{-Y_1}$.
},
the super LG model integral is given by  
\bea
\int
\left( \prod_{I=1}^{4} d y_I \right)
\frac{d x_2}{x_2^4} d \eta_2 d \chi_2 
 \; \exp \left[   
\sum_{I=1}^{4} y_I  +  
x_2 \left(1+ \eta_2 \chi_2\right)
+ \frac{e^{t } y_1 y_2 y_3 y_4}{x_2^3} \right].
\nonu
\eea
By redefining 
\bea
\widetilde{y}_1 =y_1, \qquad
\widetilde{y}_i \equiv \frac{y_i}{x_2}, \qquad 
i=2,3,4
\nonu
\eea 
in order to
make the super 
LG effective superpotential in the exponent to be 
a polynomial form(for Calabi-Yau supermanifold),
one can reexpress it as 
\bea
\int 
\left( \prod_{I=1}^{4} d \widetilde{y}_I \right)
\left( \frac{d x_2}{x_2} \right) d \eta_2 d \chi_2
 \; \exp \left[   \widetilde{y}_1
+\sum_{i=2}^{4} x_2 \widetilde{y}_i  +  
x_2\left(1+ \eta_2 \chi_2 \right) 
+ e^t \prod_{I=1}^{4} \widetilde{y}_I    \right].
\nonu
\eea
Note that $\widetilde{y}_1$ is a Lagrange multiplier
whose equation of motion is given by $1+e^t \prod_{I=2}^4 
\widetilde{y}_I=0$.
In order to absorb the nontrivial measure $1/x_2$ for $x_2$,
let us introduce the 
two additional chiral bosonic superfields 
$u$ and $v$ taking values in ${\bf C}$ 
through a relation
\bea
\int du dv e^{uv x_2} = \frac{1}{x_2}
\nonu
\eea
enforcing $x_2$ to become a Lagrange multiplier 
due to the algebraic constraint.
%
By performing the $x_2$ and $\widetilde{y}_1$-integrations,
the LG period turns out to be 
\bea
\int 
\left( \prod_{i=2}^{4} d \widetilde{y}_i \right)
 du dv  d \eta_2 d \chi_2
\; \de\left( 1- uv + \sum_{i=2}^{4}  \widetilde{y}_i+
\eta_2 \chi_2  \right)
 \de \left(  1
+ e^t \prod_{i=2}^{4} \widetilde{y}_i    \right).
\nonu
\eea
Finally the $\widetilde{y}_4$-integration(with the replacement of
$t \rightarrow t +i\pi$) gives
the following integral with 3 bosonic and 2 fermionic 
degrees of freedom(4 bosonic coordinates has one delta function
constraint)
\bea
\int 
d \widetilde{y}_2 d \widetilde{y}_3
 du dv  d \eta_2 d \chi_2
\; \de\left( 1- uv + \widetilde{y}_2 +\widetilde{y}_3 
+ \frac{e^{-\widetilde{t}}}{\widetilde{y}_2 \widetilde{y}_3} +
\eta_2 \chi_2  \right).
\label{expression}
\eea
The delta function inside the integral contains the information
on the geometry of the mirror Calabi-Yau supermanifold.

Then the super LG mirror of 
${\bf WCP }^{3|2}(1,1,1,1|1,3)$, 
by integrating out the dual fields
corresponding to the fermion of weight 1,
can be regarded as a super 
Calabi-Yau hypersurface
characterized by
\bea
 1- uv + \widetilde{y}_2 +\widetilde{y}_3 
+ \frac{e^{-\widetilde{t}}}{\widetilde{y}_2 \widetilde{y}_3} +
\eta_2 \chi_2=0
\label{hypersurface}
\eea
where $\widetilde{y}_2$ and $\widetilde{y}_3$ take values in 
${\bf C}^{\ast}$ and $u$ and $v$ are variables in ${\bf C}$.
One can split this as 
\bea
g(\widetilde{y}_2, \widetilde{y}_3)= u v - \eta_2 \chi_2, 
\qquad \mbox{and} \qquad
g(\widetilde{y}_2, \widetilde{y}_3) = 1 +
\widetilde{y}_2 +\widetilde{y}_3 
+ \frac{e^{-\widetilde{t}}}{\widetilde{y}_2 \widetilde{y}_3}.
\nonu
\eea
The three critical points of 
$g(\widetilde{y}_2, \widetilde{y}_3)$
are given by $\widetilde{y}_i=\omega 
e^{-\frac{\widetilde{t}}{3}}$ where $\omega$ is a 3rd root
of unity which yields the critical values in the 
$g(\widetilde{y}_2, \widetilde{y}_3)$ hypersurface
on three points on a tiny circle of radius 
$|e^{-\frac{\widetilde{t}}{3}}|$ near 
$g(\widetilde{y}_2, \widetilde{y}_3)=1$.
This noncompact mirror Calabi-Yau has complex superdimension
$(3|2)$ in toric supermanifold
which is exactly the same as the dimension of the original
${\bf WCP }^{3|2}(1,1,1,1|1,3)$.
Both supermanifolds have the same superdimension 1, as required
by mirror symmetry because the superdimension determines
the $\hat{c}$ in the current algebra. 

At $\eta_2=0=\chi_2$ patch, 
this hypersurface (\ref{hypersurface}) 
in toric variety is exactly the noncompact 
bosonic Calabi-Yau threefold which is mirror to 
another noncompact bosonic 
Calabi-Yau threefold(there exists an obvious run-away direction of
the superpotential), the line bundle ${\cal O}(-3)$ 
over ${\bf CP}^2$ that is realized by 
a linear sigma model description in terms of a single $U(1)$
gauge theory with charges of the matter fields $(-3,1,1,1)$ without 
superpotential term \cite{HIV}
\footnote{The LG superpotential of the mirror theory is given by
$W=x_0 + x_1 + x_2 +e^{-t} \frac{x_0^3}{x_1 x_2}$ where 
$x_i=e^{-Y_i}$. Here $Y_0$ is the dual field to the charge $-3$ 
matter field and $Y_i(i=1,2,3)$ to the charge 1 matter fields. 
At $e^{t}=-27$, the singularity appears $\frac{\pa W}{\pa x_i}=0$. 
In the last term of the superpotential we replaced 
$Y_3$ with $3Y_0-Y_1-Y_2+t$ using a delta 
function constraint. Then it is easy to see \cite{HIV} 
that the noncompact
Calabi-Yau threefold defined by this LG superpotential is 
equivalent to another noncompact Calabi-Yau threefold defined by 
the above (\ref{hypersurface}) with $\eta_2=0=\chi_2$.  
\label{foot} }. 
The field of $-3$ charge parametrizes the complex direction of the 
fiber and the fields with 1 charge correspond to 
span the base ${\bf CP}^2$.
In this sense, 
the original Calabi-Yau supermanifold    
${\bf WCP }^{3|2}(1,1,1,1|1,3)$ contains  
noncompact bosonic Calabi-Yau threefold:
the line bundle ${\cal O}(-3)$ 
over ${\bf CP}^2$.

The holomorphic volume form can be viewed as
\bea
\Omega = \frac{d \widetilde{y}_2
d \widetilde{y}_3 du dv d \eta_2 d \chi_2}{df},
\qquad f \equiv
1- uv + \widetilde{y}_2 +\widetilde{y}_3 
+ \frac{e^{-\widetilde{t}}}{\widetilde{y}_2 \widetilde{y}_3} +
\eta_2 \chi_2 =0 
\nonu
\eea
and further $v$-integration on (\ref{expression}) gives
the period of the holomorphic volume form
$
\int  \Omega
$. 
For better understanding the geometry of the mirror,
by rescaling the fields \cite{HIV},
\bea
\widetilde{y}_2 \rightarrow e^{-\frac{\widetilde{t}}{3}} 
\widetilde{y}_2,  
\quad 
\widetilde{y}_3 \rightarrow e^{-\frac{\widetilde{t}}{3}} 
\widetilde{y}_3,
\quad
u \rightarrow e^{-\frac{\widetilde{t}}{6}} u,
\quad
v \rightarrow e^{-\frac{\widetilde{t}}{6}} v,   
\quad
\eta_2 \rightarrow e^{-\frac{\widetilde{t}}{6}} \eta, 
\quad
\chi_2 \rightarrow e^{-\frac{\widetilde{t}}{6}} \chi
\nonu
\eea 
the defining equation (\ref{hypersurface}) will be
\bea
\widetilde{y}_2^2 \widetilde{y}_3 + 
\widetilde{y}_2 \widetilde{y}_3^2  + \widetilde{y}_2 
\widetilde{y}_3
z +1=0, 
\qquad 
z = e^{\frac{\widetilde{t}}{3}} - u v + \eta 
\chi,
\nonu
\eea 
and finally this 
\footnote{
One can introduce an extra 
variable $x$ in order to make the equation homogenize:
$
\widetilde{y}_2^2 \widetilde{y}_3 + 
\widetilde{y}_2 \widetilde{y}_3^2  + x \widetilde{y}_2 
\widetilde{y}_3
z +x^3=0$.
Then we take the coordinate transformation given by 
\cite{HIV}, 
$
\widetilde{y}_2 = 
y-\left( \frac{z}{2} \right) x +\frac{1}{2}$ and $
 \widetilde{y}_3 = -y -\left( \frac{z}{2} \right) x + 
\frac{1}{2}$. } 
can be written in Weierstrass form
\bea
y^2 = x^3 + \left( \frac{z}{2}\right)^2 x^2 -
\left(\frac{z}{2}\right) x +\frac{1}{4}, \qquad
z- e^{\frac{\widetilde{t}}{3}}= -uv +\eta \chi.
\label{Equa}
\eea
The first equation defines an elliptic
fibration over the complex plane with coordinate $z$ and the 
second equation describes a ${\bf C}^{\ast}$-fibration over the
$(z,\eta,\chi)$-surface because 
for fixed values of $(z,\eta,\chi)$, from a relation $uv =
\mbox{const}$, $v$ can be written as 
$v=\mbox{const}/u$ and $u$ can be any nonzero 
complex value. The general fiber is ${\bf C}^{\ast}$. 
At $z- e^{\frac{\widetilde{t}}{3}}=\eta \chi$, the 
${\bf C}^{\ast}$-fibration degenerates when its nontrivial
${\bf S}^1$ shrinks.
One can also interpret this as ${\bf C}^2$-fibration over 
$(z,u)$-surface since for fixed values of $(z,u)$, one can 
express $v$ in terms of $\eta,\chi$ that can be any complex values.
Recall that $u$ and $v$ are bosonic 
superfields while $\eta$ and $\chi$ are {\it fermionic} 
superfields.

Therefore, 
the super LG mirror of 
${\bf WCP }^{3|2}(1,1,1,1|1,3)$, by integrating out the dual 
superfields
corresponding to the fermion of weight 1, can be regarded as
an elliptic fibration over $z$-plane with the second equation of 
(\ref{Equa}).

\subsection{
Integration over $X_2$}
\label{X2}

\indent

Let us consider the case where we integrate out $X_2$
instead of $X_1$.
In this case, the delta function constraint 
will provide
\bea
X_2=-\frac{1}{3} X_1  -\frac{t}{3}+
\frac{1}{3} \sum_{I=1}^{4} Y_I. 
\nonu
\eea
By computing the integrations over $X_2, \eta_2$ and $\chi_2$
as we have done before,
one gets the following B-model integral with 5 bosons and 2
fermions
\bea 
\int \prod_{I=1}^{4} d Y_I 
d X_1 d \eta_1 d \chi_1
\;  e^{\frac{1}{3} X_1  -
\frac{1}{3} \sum_{I=1}^{4} Y_I} 
 \exp \left[   
\sum_{I=1}^{4} e^{-Y_I}  +  
e^{-X_1} \left(1+ \eta_1 \chi_1\right)
+ e^{ \frac{1}{3} X_1  +\frac{t}{3}-
\frac{1}{3} \sum_{I=1}^{4} Y_I
 }  \right].
\nonu
\eea
Now let us introduce the new $\bf C$-valued fields 
$x_1$ and $y_I$ by realizing the measure factors:
\bea
e^{-\frac{X_1}{3}} \equiv    x_1, \qquad e^{-\frac{Y_I}{3}} 
\equiv y_I.
\nonu
\eea

Then in terms of these new fields the super 
LG model is  given by
\bea
\int
\left( \prod_{I=1}^{4} d y_I \right)
\left( \frac{d x_1}{x_1^2} \right) d \eta_1 d \chi_1 
 \exp \left[   
\sum_{I=1}^{4} y_I^3  +  
x_1^3 \left(1+ \eta_1 \chi_1\right)
+ \frac{e^{\frac{t}{3} } y_1 
y_2 y_3 y_4}
{x_1} \right].
\nonu
\eea
By redefining 
\bea
\widetilde{y}_1 \equiv \frac{y_1}{x_1}, \qquad
\widetilde{y}_i =y_i, \qquad i=2,3,4
\nonu
\eea 
in order to
make the super 
LG superpotential in the exponent to be 
a polynomial form,
one arrives at
\bea
\int 
\left( \prod_{I=1}^{4} d \widetilde{y}_I \right)
\left( \frac{d x_1}{x_1} \right) d \eta_1 d \chi_1
 \exp \left[  x_1^3 \widetilde{y}_1^3
+\sum_{i=2}^{4}  \widetilde{y}_i^3  +  
x_1^3 \left(1+ \eta_1 \chi_1 \right) 
+ e^{\frac{t}{3}} \prod_{I=1}^{4} \widetilde{y}_I    \right].
\nonu
\eea
By using the relation 
$
x_1^3 \equiv \widetilde{x}_1
$
one can express this as
\bea
\int 
\left( \prod_{I=1}^{4} d \widetilde{y}_I \right)
\left( \frac{d \widetilde{x}_1}{\widetilde{x}_1} 
\right) d \eta_1 d \chi_1
 \exp \left[  \widetilde{x}_1 \widetilde{y}_1^3
+\sum_{i=2}^{4}  \widetilde{y}_i^3  +  
\widetilde{x}_1 \left(1+ \eta_1 \chi_1 \right) 
+ e^{\frac{t}{3}} \prod_{I=1}^{4} \widetilde{y}_I    \right].
\nonu
\eea 

In order to absorb the nontrivial measure $1/\widetilde{x}_1$ 
for $\widetilde{x}_1$ as we have done before,
let us introduce the 
two additional chiral superfields $u$ and $v$ through 
$
\int du dv e^{uv \widetilde{x}_1} = \frac{1}{\widetilde{x}_1}
$ allowing  $\widetilde{x}_1$ to become a Lagrange multiplier.
Now $\widetilde{x}_1$-integration gives
\bea
\int 
\left( \prod_{I=1}^{4} d \widetilde{y}_I \right)
 du dv  d \eta_1 d \chi_1
\; \de \left( \widetilde{y}_1^3 + 1+ \eta_1 \chi_1 -u v \right) 
\exp \left[ 
\sum_{i=2}^{4}  \widetilde{y}_i^3  
+ e^{\frac{t}{3}} \prod_{I=1}^{4} \widetilde{y}_I        \right].
\label{expr}
\eea
Thus we have obtained a super submanifold 
defined by
$
\widetilde{y}_1^3 + 1+ \eta_1 \chi_1 -u v =0
$
and the expression (\ref{expr}) is identical to the period
of super LG model of submanifold with superpotential
\bea
W = \sum_{i=2}^{4}  \widetilde{y}_i^3  
+ e^{\frac{t}{3}} \prod_{I=1}^{4} \widetilde{y}_I=
 \sum_{i=2}^{4}  \widetilde{y}_i^3  
+ \left(e^{\frac{t}{3}} \widetilde{y}_1 \right) 
\prod_{i=2}^{4} \widetilde{y}_i.   
\label{superp}
\eea
How do we interpret this?
For gauged linear sigma model \cite{Witten1993} 
where a single $U(1)$ theory  
with charged matter fields with charges given by
$(-n,1,1,1)$ in complex
dimension 3, the 
$n=3$ case is conformal and corresponds to 
the ${\cal O}(-3)$ geometry over ${\bf CP}^2$ or its orbifold limit
when $t \rightarrow -\infty$ given by ${\bf C}^3/{\bf Z}^3$. 
In this limit the superpotential has a simple form
without the second term of (\ref{superp}) \cite{Vafa2001,HoriVafa}.
In the present case, the dual field corresponding to 
the charge $-3$ is replaced by other dual fields through the
delta function constraint, contrary to the previous case explained in
the footnote \ref{foot} where one of the matter fields with 
charge 1 was replaced.
Therefore when we integrated out a dual superfield $X_1$ 
corresponding to a fermionic 
superfield of weight 1 as in previous subsection \ref{X1},
this led to  the ${\cal O}(-3)$ geometry over ${\bf CP}^2$
with the replacement of charge 1 in a linear sigma model.
On the other hand, when we integrate out a dual superfield
$X_2$ corresponding to a fermionic superfield of weight 3 in this
subsection, it produces the same Calabi-Yau threefold with the 
replacement of charge 3.   
The mirror description for linear sigma model 
is exactly given by the  
$\widetilde{y}_1=1$ patch of (\ref{superp}) modulo 
${\bf Z}_3 \times {\bf Z}_3$ which is the maximal group preserving 
all the monomials and ${\bf Z}_3$ acts as 3rd roots of unity on 
each field preserving all the monomials.

By the following redefinitions
\bea
\widetilde{y}_1 = \hat{y}_1, \qquad
\widetilde{y}_2^3 = \hat{y}_2, \qquad
\widetilde{y}_3 = \widetilde{y}_2 \hat{y}_3, \qquad
\widetilde{y}_4 = \widetilde{y}_2 \hat{y}_4
\nonu
\eea
and after $\hat{y}_2$-integration,
the super LG model will lead to the following integral with 
3 bosons and 2 fermions(there are two delta functions)
\bea
\int 
d \hat{y}_1 d \hat{y}_3  d \hat{y}_4
 du dv  d \eta_1 d \chi_1
\; \de \left( \hat{y}_1^3 + 1+ \eta_1 \chi_1 -u v \right) 
\de \left( 
1+   \hat{y}_3^3 +\hat{y}_4^3  
+ e^{\frac{t}{3}} \hat{y}_1 \hat{y}_3 \hat{y}_4 \right).
\nonu
\eea
Then the mirror of 
 ${\bf WCP }^{3|2}(1,1,1,1|1,3)$ by integrating out the dual fields
 corresponding to the fermion of weight 3 can be regarded as a super 
Calabi-Yau hypersurface
characterized by
\bea
\hat{y}_1^3 + 1+ \eta_1 \chi_1 -u v =0, \qquad
1+   \hat{y}_3^3 +\hat{y}_4^3  
+ e^{\frac{t}{3}} \hat{y}_1 \hat{y}_3 \hat{y}_4 =0.
\label{twoeq}
\eea
It is more convenient to introduce an extra variable 
$\hat{y}_0$ in the second equation of (\ref{twoeq}):
\bea
\hat{y}_0^3+   \hat{y}_3^3 +\hat{y}_4^3  
+ \left( e^{\frac{t}{3}} \hat{y}_1 \right) \hat{y}_0 \hat{y}_3 \hat{y}_4=0. 
\label{eq}
\eea
Then the equation (\ref{eq}) is invariant under 
\bea
\hat{y}_0 \rightarrow \ga_0 \hat{y}_0, \quad 
\hat{y}_3 \rightarrow \ga_3 \hat{y}_3, \quad
\hat{y}_4 \rightarrow \ga_4 \hat{y}_4, \quad
\hat{y}_1 \rightarrow \ga_1 \hat{y}_1, \quad
\ga_i^3 =1(i=0,3,4), 
\quad \ga_0 \ga_1 \ga_3 \ga_4 =1.
\nonu
\eea
This is the equation for a Calabi-Yau 
hypersurface in ${\bf CP}^2$ with the parameter 
$e^{\frac{t}{3}} \hat{y}_1$:
the degree 3 Calabi-Yau hypersurface 
in  ${\bf CP}^2$ fibered over $\bf C$ \cite{HoriVafa}.
The first equation of (\ref{twoeq}) is a 
${\bf C}^{\ast}$-fibration over $(\hat{y}_1, 
\eta_1, \chi_1)$-surface because for fixed these values 
the relation $uv=\mbox{const}$ determines $v$ in terms of 
nonzero complex value $u$ which provides the fiber 
${\bf C}^{\ast}$.
The second equation of (\ref{twoeq}) is a
$\hat{y}_0=1$ patch of (\ref{eq}). 
Of course, one can reduce to a single equation
by susbstituting $\hat{y}_1$ from the first equation of 
(\ref{twoeq}) into the second equation of (\ref{twoeq}), but 
it is not clear whether 
this has any simple geometric interpretation or not.

Therefore, 
the super LG mirror of 
${\bf WCP }^{3|2}(1,1,1,1|1,3)$, by integrating out 
the dual superfields
corresponding to the fermion of weight 3, can be regarded as
some patch of a degree 3 Calabi-Yau hypersurface 
in ${\bf CP}^2$ fibered 
by ${\bf C}$  with the first equation of 
(\ref{twoeq}).

\section{Concluding remarks}
\setcounter{equation}{0}

\indent

In this paper, 
we have found that
the super Landau-Ginzburg B-model mirrors of 
Calabi-Yau super manifold ${\bf WCP}^{3|2}(1,1,1,1|1,3)$
can be described by a super Calabi-Yau hypersurface 
(\ref{hypersurface})(or (\ref{Equa})) or
(\ref{twoeq})(or (\ref{superp})) 
depending on which  dual 
superfields we take.
As an bosonic submanifold, the original 
${\bf WCP}^{3|2}(1,1,1,1|1,3)$ contains
a noncompact Calabi-Yau threefold, a line bundle 
${\cal O}(-3)$ over ${\bf CP}^2$.
The two dual {\it fermionic} superfields 
enter into either an elliptic fibration or 
some patch of cubic Calabi-Yau hypersurface in ${\bf CP}^2$ 
fibered over ${\bf C}$  differently
because the original weighted projective superspace
${\bf WCP}^{3|2}(1,1,1,1|1,3)$ possesses different weights in 
the  {\it fermionic} superfields.

Although other bosonic complex manifold 
${\bf CP}^{M-1}$($M \neq 4$) is not related to 
four-dimensional Minkowski spacetime by the Penrose 
transform \cite{Witten}, 
one  also 
apply for the mirror of higher(only for bosonic dimension) 
dimensional 
weighted projective space 
${\bf WCP}^{5|2}(1,1,\cdots, 1|1,5)$ which has a linear sigma model
in terms of six bosonic homogeneous coordinates 
$Z^I$ of weight 1 and two fermionic coordinates 
$\psi$ and $\chi$ of weights 1 and 5, respectively.
This is also Calai-Yau supermanifold.
Under the T-duality, the corrresponding super LG mirror of
 ${\bf WCP}^{5|2}(1,1,\cdots,1|1,5)$ is written as
a path integral with eight bosonic and four 
fermionic dual variables with one single delta function 
and the superpotential has an extra two terms due to the 
extra two bosonic homogeneous coordinates, compared with 
${\bf WCP}^{3|2}$. 
Following the procedures, i) computation for the integrations on
$\eta_1, \chi_1$ and $X_1$, ii) change to the right variables,
iii) absorbing the nontrivial factor $1/x_2$, iv) integration for
Lagrange multipliers, and v) using a delta function, 
as we have done in subsection 
\ref{X1}, 
one gets
\bea
\int 
\left(\prod_{i=2}^{5} d \widetilde{y}_i \right) 
du dv  d \eta_2 d \chi_2
\; \de\left( 1- uv + \sum_{i=2}^{5} \widetilde{y}_i
+ \frac{e^{-\widetilde{t}}}{\prod_{i=2}^{5} \widetilde{y}_i} +
\eta_2 \chi_2  \right).
\label{equation5-1}
\eea
At $\eta_2=0=\chi_2$ patch, 
the corresponding  
hypersurface in toric variety is exactly the noncompact 
bosonic Calabi-Yau threefold which is equivalent to 
another noncompact bosonic 
Calabi-Yau threefold, the line bundle ${\cal O}(-5)$ 
over ${\bf CP}^4$. This is realized by 
a linear sigma model description in terms of a single $U(1)$
gauge theory with charges of the matter fields $(-5,1,1,1,1,1)$ 
without 
superpotential term \cite{CG,HIV}. 
The field of $-5$ charge corresponds to the 
fiber coordinate and the fields with 1 charge correspond to 
span the base ${\bf CP}^4$. For negative $\mbox{Re}\; t$, 
the space of fields is ${\bf C}^5/{\bf Z}_5$. The blow up of the 
origin in ${\bf C}^5/{\bf Z}_5$  is the line bundle   
 ${\cal O}(-5)$ over ${\bf CP}^4$.

On the other hand, the $X_2$-integration gives
other mirror. We can repeat the calculations by inserting the 
extra two dual bosonic variables and realizing the delta function
constraint, as 
we have done in subsection 
\ref{X2}.
One arrives at
\bea
\int 
d \hat{y}_1 
\left( \prod_{i=3}^{6} d \hat{y}_i \right)  
 du dv  d \eta_1 d \chi_1
\; \de \left( \hat{y}_1^5 + 1+ \eta_1 \chi_1 -u v \right) 
\de \left( 
1+   \sum_{i=3}^{6} \hat{y}_i^5   
+ e^{\frac{t}{5}} \hat{y}_1 \prod_{i=3}^{6} \hat{y}_i \right).
\label{equation5-2}
\eea
The second delta function, when we introduce an extra 
variable $\hat{y}_0$, provides 
the equation for the Calabi-Yau hyperspace in ${\bf CP}^4$
with the parameter $e^{\frac{t}{5}} \hat{y}_1$. That is,
a degree 5(well-known quintic) 
Calabi-Yau hypersurface in ${\bf CP}^4$ fibered over
${\bf C}$. 

Therefore, we expect that from the Calabi-Yau supermanifold 
${\bf WCP}^{M-1|2}(1,1,\cdots,1|1,M-1)$ which contains 
the line bundle   
 ${\cal O}(-(M-1))$ over ${\bf CP}^{M-2}$,
as a 
bosonic submanifold, there exist two mirror Calabi Yau 
supermanifolds for general $M$ characterized by two equations 
similar
to (\ref{equation5-1}) and (\ref{equation5-2}), by simple 
generalization(the indices for summation and product run from 
1 to $M-2$).
 
For the weighted projective superspace with a {\it single} 
fermionic superfield,
we expect, after some path integral, that the mirror we get
corresponds to bosonic Calabi-Yau hypersurface in 
ordinary weighted projective space.
One can consider the Calabi-Yau supermanifold ${\bf WCP}^{M-1|1}(1,1,
\cdots, 1|M)$ 
which has a linear sigma model 
description in terms of $M$ bosonic homogeneous 
coordinates $Z^I$ where $I=1,2, \cdots, M$ of 
weight 1 and one fermionic 
homogeneous coordinate $\psi$ of weight $M$.
Then the super LG B-model mirror of  
${\bf WCP}^{M-1|1}(1,1,
\cdots, 1|M)$ is
given by the path integral similarly \cite{AV}.
The super LG model has $(M+1)$ bosonic and 2 fermionic 
degrees of freedom. 
Let us integrate out the fermions $\eta$ and $\chi$,
solve the delta function constraint for $X$,
and introduce the new $\bf C$-valued bosonic superfield 
$y_I$ as follows:$
e^{-\frac{Y_I}{M}} \equiv y_I, 
I=1,2, \cdots, M$.
Then the super LG model is given by  
$
\int
 \prod_{I=1}^{M} d y_I 
 \; \exp \left[   
\sum_{I=1}^{M} y_I^M  
+ e^{\frac{t}{M} } \prod_{I=1}^M y_I \right]$.
%
This is exactly the 
bosonic Calabi-Yau manifold which is mirror to 
another  bosonic 
Calabi-Yau manifold, 
that is realized by 
a linear sigma model description in terms of a single $U(1)$
gauge theory with charges of the matter fields $(-M,1,1, \cdots,
1)$. 
The field of $-M$ charge corresponds to the 
fiber coordinate and the fields with 1 charge correspond to 
span the base ${\bf CP}^{M-1}$ \cite{HoriVafa}.   
The super LG yields directly the periods of the bosonic Calabi-Yau 
manifold \cite{Schwarz,AV}. Of course, the LG theory is given by
the above superpotential modded out by $({\bf Z}_M)^{M-1}$
which acts on each $y_I$ by all $M$-th roots of unity preserving 
the product $\prod_{I=1}^{M} y_I$.

We would like to list several other interesting open problems
for future directions.

$\bullet$ We have assumed the existence of 
a topological A-model on  ${\bf WCP}^{3|2}$.
It would be interesting to study
how S-dual works \cite{NV}, if there is,  between A-model and B-model on 
the same Calabi-Yau supermanifold ${\bf WCP}^{3|2}$ and 
how the extra branes in the A-model and B-model affect 
this twistorial
Calabi-Yau supermanifold.

$\bullet$ How the ${\cal N}=1$ superconformal field theory in four dimensions 
corresponding to 
a topological B-model on ${\bf WCP}^{3|2}(1,1,1,1|1,3)$ couples
to conformal supergravity? Recently, 
it was found that the action of holomorphic
Chern-Simons theory leads to some truncation of the self-dual 
${\cal N}=4$ super Yang-Mills theory \cite{PW}. 
Then it would be interesting
to find out the agreement of the twistor superfields in 
twistor-string theory with the 
physical states of ${\cal N}=4$ conformal supergravity in four 
dimensions \cite{BW}, by some truncation. Or one can 
study it from the ${\cal N}=1$ conformal supergravity 
\cite{FT} in four
dimensions directly.  
As already observed in \cite{Witten}, the topological 
B-model with target 
${\bf WCP}^{3|2}$ should preserve ${\cal N}=1$
superconformal symmetry 
$SU(4|1)$ (while in ${\cal N}=1$ super Yang-Mills theory there is
a conformal anomaly)  and has additional symmetry 
which does not exist in  ${\cal N}=1$ 
super Yang-Mills theory. 

$\bullet$ In this note, we have started with a topological A-model
on ${\bf WCP}^{3|2}$ 
and ended up with super LG B-model.    
How can we obtain super LG B-model mirror 
${\bf WCP}^{3|2}$ from some
unknown A-model? What is the correct A-model description?
It would be interesting to see this by studying our super Calabi-Yau
hypersurfaces carefully in some particular limit of $t$ and to 
find out the correct linear sigma model description 
with appropriate charge
assignments.
Since for the Calabi-Yau condition, no $U(1)$ anomaly is allowed,
the field content and $U(1)$ charge assignment should reflect this 
fact. 

$\bullet$ There exist other kinds 
of Calabi-Yau supermanifolds by giving 
different weights for the fermionic superfields: 
${\bf WCP}^{3|2}(1,1,1,1|2,2)$ and 
${\bf WCP}^{3|2}(1,1,1,1|0,4)$ \cite{PW}. Now it is straightforward 
to apply our method to these Calabi-Yau supermanifolds.
For the former,
there exists a super submanifold defined by
$1-uv + \widetilde{y}_3^2 + \widetilde{y}_4^2
+\eta_2 \chi_2=0$ togther with the period of super LG model
of submanifold with superpotential 
$
W = \sum_{i=1}^{2}  \widetilde{y}_i^2  
+ e^{\frac{t}{2}} \prod_{i=1}^{4} \widetilde{y}_i$.
Since the weights of the fermions are equal, it does not matter which 
one we integrate out.
For the latter, we can obtain a super submanifold defined by
$1+\eta_1 \chi_1=0$ togther with the period of super LG model
of submanifold with superpotential 
$
W = \sum_{i=1}^{4}  \widetilde{y}_i^4  
+ e^{\frac{t}{4}} \prod_{i=1}^{4} \widetilde{y}_i$.
There is only one mirror description because in the delta function constraint,
there is no dependence on 
the dual bosonic superfield corresponding to
fermion superfield of weight 0. 

\vspace{1cm}
\centerline{\bf Acknowledgments}
\indent

I would like to thank E. Buchbinder F. Cachazo, M. Fabinger,
S. Hellerman, S. Prem Kumar, K. Oh, 
P. Svrcek, C. Vafa and 
E. Witten for discussions. 
This research was supported by a grant in aid from the 
Monell Foundation through Institute for Advanced Study, 
by SBS Foundation, 
and by Korea Research Foundation Grant (KRF-2002-015-
CS0006).

\end{document}